\documentclass[letter,10pt]{article}

\usepackage[margin=.93in]{geometry}
\usepackage{multicol}
\usepackage{amsmath}

\makeatletter
\renewcommand*{\@fnsymbol}[1]{\ensuremath{\ifcase#1\or \dagger\or *\or \else\@ctrerr\fi}}
\makeatother
\usepackage{tikz}
\usetikzlibrary{svg.path}
\usepackage{scalerel}
\definecolor{orcidlogocol}{HTML}{A6CE39}
\tikzset{
 orcidlogo/.pic={
 \fill[orcidlogocol] svg{M256,128c0,70.7-57.3,128-128,128C57.3,256,0,198.7,0,128C0,57.3,57.3,0,128,0C198.7,0,256,57.3,256,128z};
 \fill[white] svg{M86.3,186.2H70.9V79.1h15.4v48.4V186.2z}
 svg{M108.9,79.1h41.6c39.6,0,57,28.3,57,53.6c0,27.5-21.5,53.6-56.8,53.6h-41.8V79.1z M124.3,172.4h24.5c34.9,0,42.9-26.5,42.9-39.7c0-21.5-13.7-39.7-43.7-39.7h-23.7V172.4z}
 svg{M88.7,56.8c0,5.5-4.5,10.1-10.1,10.1c-5.6,0-10.1-4.6-10.1-10.1c0-5.6,4.5-10.1,10.1-10.1C84.2,46.7,88.7,51.3,88.7,56.8z};
 }
}

\newcommand\orcid[1]{\href{https://orcid.org/#1}{\mbox{\scalerel*{
\begin{tikzpicture}[yscale=-1,transform shape]
\pic{orcidlogo};
\end{tikzpicture}
}{|}}}}

\usepackage[utf8]{inputenc}
\usepackage{hyperref}
\hypersetup{
	unicode,
	pdfauthor={Julien Barrier, Piranavan Kumaravadivel, Roshan Krishna Kumar, L. A. Ponomarenko, Na Xin, Matthew Holwill, Ciaran Mullan, Minsoo Kim, R. V. Gorbachev, M. D. Thompson, J. R. Prance, T. Taniguchi, K. Watanabe, I. V. Grigorieva, K. S. Novoselov, A. Mishchenko, V. I. Fal’ko, A. K. Geim and A. I. Berdyugin},
	pdftitle={Long-range ballistic transport of Brown-Zak fermions in graphene superlattices},
	pdfsubject={},
	pdfkeywords={},
	pdfproducer={LaTeX},
	pdfcreator={pdflatex}
}
\usepackage{textgreek}
\usepackage[labelfont=bf]{caption}

\usepackage[superscript,biblabel,nomove]{cite}
\makeatletter
\renewcommand{\@openbib@code}{%
\setlength{\parsep}{0pt}
\setlength{\itemsep}{1pt}
}
\makeatother

\title{\textbf{Long-range ballistic transport of Brown-Zak fermions in graphene superlattices}}
\author{Julien Barrier \orcid{0000-0002-6484-2157} $^{1,2}$\thanks{These authors contributed equally.}
\and Piranavan Kumaravadivel \orcid{0000-0002-9817-1697} $^{1,2}$\footnotemark[1]
\and Roshan Krishna Kumar \orcid{0000-0003-0857-4466} $^{1,2}$
\and L. A. Ponomarenko \orcid{0000-0003-1974-0642} $^{1,3}$
\and Na Xin $^{1,2}$
\and Matthew Holwill \orcid{0000-0002-8715-4402} $^{2}$
\and Ciaran Mullan \orcid{0000-0001-6319-3135} $^{1}$
\and Minsoo Kim \orcid{0000-0001-6304-6901} $^{1}$
\and R. V. Gorbachev \orcid{0000-0003-3604-5617} $^{1,2}$
\and M. D. Thompson \orcid{0000-0002-8551-4806} $^{2,3}$
\and J. R. Prance \orcid{0000-0001-5009-383X} $^{2,3}$
\and T. Taniguchi$^{4}$
\and K. Watanabe \orcid{0000-0003-3701-8119} $^{4}$
\and I. V. Grigorieva \orcid{0000-0001-5991-7778} $^{1,2}$
\and K. S. Novoselov \orcid{0000-0003-4972-5371} $^{1,2}$
\and A. Mishchenko \orcid{0000-0002-0427-5664} $^{1,2}$
\and V. I. Fal’ko \orcid{0000-0003-0828-0310} $^{1,2}$
\and A.K. Geim \orcid{0000-0003-2861-8331} $^{1,2}$\thanks{Correspondance for this work shold be addressed to A.I. Berdyugin \texttt{alexey.berdyugin@manchester.ac.uk} and A.K. Geim \texttt{geim@manchester.ac.uk}}
\and A.I. Berdyugin \orcid{0000-0002-7537-6227} $^{1,2}$ \footnotemark[2]}

\date{\small
	$^1$ Department of Physics and Astronomy, University of Manchester, Manchester M13~9PL, UK\\%
	$^2$ National Graphene Institute, University of Manchester, Manchester M13~9PL, UK\\%
	$^3$ Department of Physics, University of Lancaster, Lancaster LA1~4YW, UK\\%
	$^4$ National Institute for Materials Science, Ibaraki 305-0044, Japan\\[2ex]%
}

\begin{document}
\maketitle
	
\begin{abstract}
	In quantizing magnetic fields, graphene superlattices exhibit a complex fractal spectrum often referred to as the Hofstadter butterfly. It can be viewed as a collection of Landau levels that arise from quantization of Brown-Zak minibands recurring at rational ($p/q$) fractions of the magnetic flux quantum per superlattice unit cell. Here we show that, in graphene-on-boron-nitride superlattices, Brown-Zak fermions can exhibit mobilities above 10$^6$ cm$^2$V$^{-1}$s$^{-1}$ and the mean free path exceeding several micrometers. The exceptional quality of our devices allows us to show that Brown-Zak minibands are $4q$ times degenerate and all the degeneracies (spin, valley and mini-valley) can be lifted by exchange interactions below 1K. We also found negative bend resistance at $1/q$ fractions for electrical probes placed as far as several micrometers apart. The latter observation highlights the fact that Brown-Zak fermions are Bloch quasiparticles propagating in high fields along straight trajectories, just like electrons in zero field.\\[3ex]
\end{abstract}

\begin{multicols}{2}
\section*{Introduction}
	
Van der Waals assembly offers a possibility to create materials by stacking atomically-thin layers of different crystals~\cite{geim2013,novoselov2016,yankowitz2019}.
One of the simplest and most studied van der Waals heterostructures is graphene encapsulated between two hexagonal boron nitride (hBN) crystals. The encapsulation protects graphene from extrinsic disorder~\cite{dean2010,mayorov2011}, allowing ultra-high electronic quality and micrometer-scale ballistic transport often limited only by edge scattering~\cite{wang2013,kumaravadivel2019}. A special case of encapsulated graphene heterostructures is graphene superlattices where crystallographic axes of graphene and hBN are intentionally aligned. A small (1.8\%) mismatch between graphene and hBN crystal lattices results in a periodic moiré potential acting on charge carriers in graphene and leading to the formation of electronic minibands\cite{geim2013,novoselov2016,yankowitz2019,chen2020,wallbank2013,yankowitz2012,ponomarenko2013,dean2013,hunt2013,yu2014,diez2014,wang2015,spanton2018}. 

A relatively large ($\approx$14~nm) periodicity of graphene-on-hBN superlattices has also made it possible to study the regime of Hofstadter butterflies, which requires the magnetic flux $\phi$ per superlattice unit cell to be comparable to the flux quantum $\phi_0$ in experimentally-accessible magnetic fields $B$. At $B = B_{p/q}$ corresponding to $\phi=\phi_0 p/q$, where $p$ and $q$ are integer, the translational symmetry of the electronic system is restored (despite the presence of a quantizing magnetic field) and the superlattice’s electronic spectrum can again be described in terms of Bloch states~\cite{brown1964,zak1964,azbel1964,wannier1978,hofstadter1976,streda1982,rhim2012,rammal1985,delplace2010,chen2014,krishnakumar2017,krishnakumar2018}, just like in $B = 0$. These high-field Bloch states are characterized by their own miniband spectra~\cite{rammal1985,delplace2010,chen2014,krishnakumar2017,krishnakumar2018} different from the zero-$B$ spectrum of graphene-on-hBN superlattices. The associated quasiparticles are referred to as Brown-Zak (BZ) fermions. According to the group-theory analysis, an electronic spectrum for each realization of BZ fermions should have an additional $q$-fold degeneracy~\cite{brown1964,zak1964,azbel1964,wannier1978,hofstadter1976,streda1982,rhim2012,rammal1985,delplace2010,chen2014,tknn1982,landau1980} (that is, contains $q$ equivalent mini-valleys). This degeneracy is additional to the 4-fold spin and valley degeneracy of graphene’s original spectrum. Importantly, BZ fermions are Bloch quasiparticles, like electrons in solids or Dirac fermions, and, at $B = B_{p/q}$, they move through the superlattice as if the applied field is zero. Away from these exact values, BZ fermions experience an effective magnetic field $B_{eff}$ = $B - B_{p/q}$ (refs~\cite{brown1964,zak1964,azbel1964,wannier1978,hofstadter1976,streda1982,rhim2012,rammal1985,delplace2010,chen2014,tknn1982,landau1980}).

\section*{Results}
\subsection*{Experimental devices and measurement setup}
We report the electronic properties of BZ fermions with different $p$ and $q$ using high-quality graphene superlattices. These devices were fabricated using the standard dry transfer procedures, where the studied graphene crystal was carefully aligned with one of the encapsulating hBN crystals using the crystallographic edges\cite{ponomarenko2013}. The alignment was verified by Raman spectroscopy\cite{woods2014} prior to encapsulation with the second hBN crystal. The latter was intentionally misaligned to avoid competing moiré patterns\cite{wang2019a,wang2019b,finney2019}. The assembled stacks were placed on an oxidized Si wafer, which allowed us to apply the back-gate voltage $V_g$ to control the carrier density $n$. We studied six devices that were shaped into the multiterminal Hall bar geometry and had the main channel widths $W$ ranging from 2 to 17~{\textmu}m (see Fig. 1a and Supplementary Note 1). The devices were first characterized by measuring their longitudinal resistivity $\rho_{xx}$ in zero $B$ and Hall resistivity $\rho_{xy}$ in small non-quantising $B$ below 0.1~T. The latter enabled us to find the $n(V_g)$ dependences, except for gate voltages close to the neutrality points (NPs) and van Hove singularities (vHS), where $\rho_{xy}$ reversed its sign and could no longer be described by the standard dependence $\rho_{xy} = B/ne$ ($e$ is the electron charge). All our devices exhibited very high carrier mobilities $\mu = (\rho_{xx}ne)^{-1}$ of the order of 10$^6$ cm$^2$ V$^{-1}$ s$^{-1}$, which were still somewhat reduced by edge scattering because of finite W (Fig. 1b). We corroborated the high quality of our devices using transverse magnetic focusing measurements (Supplementary Note 2). In the main text, we focus on two large-width devices (D1 and D2) exhibiting highest $\mu$.

\begin{figure*}
    \centering
    \includegraphics[width=.95\textwidth]{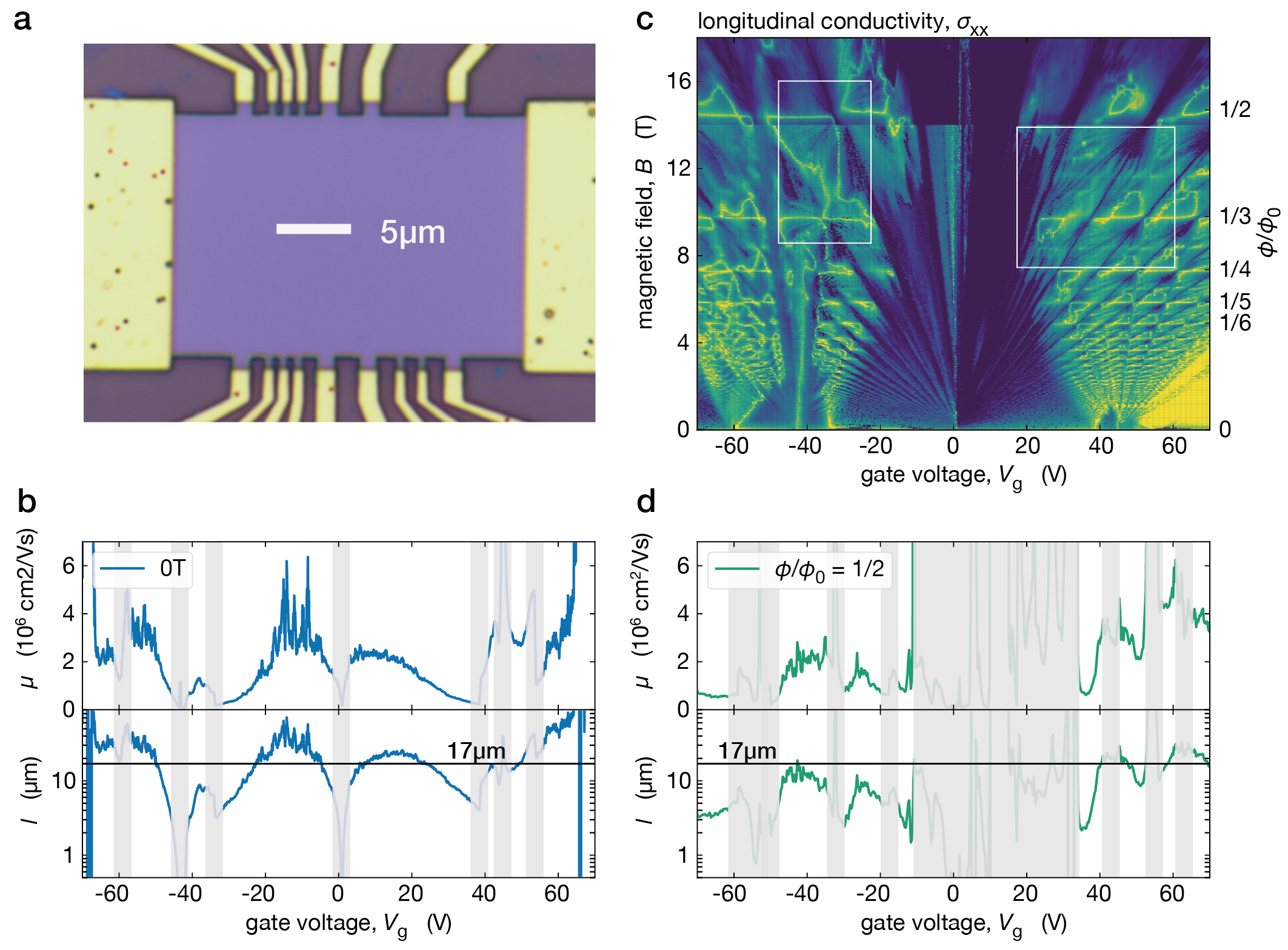}
    \caption{\textbf{High-quality graphene superlattices and their transport properties. a,} Optical micrograph of one of our devices (D1; twist angle $\theta$ between graphene and hBN of about 0.4$^{\circ}$). The Hall bar is seen in violet with golden electrical contacts. \textbf{b,} Mobility and mean free path for D1 measured at zero $B$ and 10~mK. Semitransparent vertical strips indicate the doping regions around NPs and vHS where $n$ could not be extracted directly from Hall measurements and charge inhomogeneity also plays a role. To calculate $\mu$ and $l$ within the shaded regions, we assumed a constant gate capacitance and linearly extrapolated the $n(V_g)$ dependences found sufficiently far from NPs and vHS (Supplementary Note 3). The noisy behavior at large values of $\mu$ and $l$ arises from $\rho_{xx}$ becoming small ($\approx$1~\textOmega, about 4 orders of magnitude smaller than that at the NPs). The horizontal black line indicates the device width $W$. \textbf{c,} $\sigma_{xx}(V_g, B)$ measured by sweeping $V_g$ and varying $B$ in small steps of 40~mT. $T$ = 10 and 250~mK below and above 14~T, respectively. Indigo-to-yellow colors: Log scale truncated between 38~nS and 16~mS for $B <$ 14~T and between 4~nS and 0.4~mS above 14~T. White rectangles: these regions are shown in finer detail in Figs. 3 and 4. \textbf{d,} Same as in panel b but for $\phi/\phi_0 = 1/2$ ($B \approx$ 15~T); $T$ = 250~mK. In addition to NPs and vHS, the grey strips also cover a wide region of the quantum Hall regime ($|V_g| <$ 20~V), which is dominated by large cyclotron gaps in the main graphene spectrum. The transport data used to calculate $\mu$ and $l$ in panel (b) and (d) are shown in Supplementary Note 3.}
    \label{fig:fig1}
\end{figure*}

Fig. 1c shows a map of the longitudinal conductivity $\sigma_{xx}=\rho_{xx} / (\rho_{xx}^2+\rho_{xy}^2)$ as a function of $V_g$ and $B$ in fields up to 18~T. The main features on such maps have been well understood in terms of the Hofstadter spectrum for Dirac fermions in moiré superlattices~\cite{ponomarenko2013,dean2013,hunt2013,yu2014,diez2014,wang2015,spanton2018,hofstadter1976}. One can see numerous Landau levels (LLs) fanning out from the main and secondary NPs (miniband edges) which are located at $V_g$ near 0 and $\pm$45~V, respectively. There are also pronounced funneling features near van Hove singularities at $V_g \approx$ -60, -35, +40 and +55 V in both hole ($-$) and electron ($+$) parts of the spectrum. Another important attribute of the map is horizontal yellow streaks that occur at $\phi= \phi_0 p/q$ (Fig. 1c). If temperature $T$ is increased above 100~K so that Landau quantization is strongly suppressed, the horizontal streaks become the only dominant feature on such transport maps~\cite{krishnakumar2017,krishnakumar2018}. In the high-T regime, the streaks represent oscillations in both $\rho_{xx} (B)$ and $\rho_{xy} (B)$ at a constant $V_g$. Their $1/B$ frequency is independent of $n$, and they were named BZ oscillations28. The horizontal streaks are seen in Fig. 1c correspond to maxima in $\sigma_{xx}$ and zeros in $\rho_{xy}$ and, as explained in the introduction, reflect the recovery of translational symmetry ($p$ flux quanta penetrate through $q$ superlattice unit cells) and the emergence of Bloch states experiencing zero $B_{eff}$. Along each horizontal streak, one can find numerous NPs and vHS, which reflect different realizations of BZ fermions at each $B_{p/q}$. Landau mini-fans radiate from these NPs in both directions along the $B$ axis (Fig. 1c), representing Landau quantization of BZ electronic spectra by non-zero $B_{eff}$ (see below)~\cite{brown1964,zak1964,azbel1964,wannier1978,hofstadter1976,streda1982,rhim2012,rammal1985,delplace2010,chen2014,krishnakumar2017,krishnakumar2018}. Another notable feature of the shown $\sigma_{xx}$ map is a repetitive triangular-like pattern seen most clearly between 2 and 12~T, especially for positive $V_g$. The yellow triangles are made of horizontal streaks at zero $B_{eff}$, vertical streaks emerging from NPs for BZ fermions and slanted streaks originating from vHS.

\subsection*{Ballistic transport}

The transport properties of BZ fermions can be analyzed in the same way as in Fig. 1b for Dirac fermions. The results are plotted in Fig. 1d for the case of $\phi/\phi_0 = 1/2$ and show that, away from NPs and vHS, BZ fermions in our devices exhibit $\mu$ reaching a few 10$^6$ cm$^2$ V$^{-1}$ s$^{-1}$. This is comparable to $\mu$ of Dirac fermions in zero $B$. Another important characteristic of charge carrier transport is the mean free path $l$. It can be evaluated for both Dirac and BZ fermions, using the standard expression $\sigma_{xx}=g e^2/h((k_F l)/2)$ where $h$ is the Planck constant, $k_F$ is the Fermi momentum and $g$ is the degeneracy. In zero $B$, $g = 4$ because of the spin and valley degeneracy of Dirac fermions. BZ fermions are expected to have an additional, mini-valley degeneracy~\cite{brown1964,zak1964,azbel1964,wannier1978,hofstadter1976,streda1982,rhim2012,rammal1985,delplace2010,chen2014,tknn1982,landau1980}, which is equal to $q$ (that is, $g = 8$ for the case of Fig. 1d). Using these $g$, we calculated $l$ as shown in the bottom panels of Figs. 1b and 1d. The mean free path in zero $B$ reached $>$ 20~{\textmu}m, implying that ballistic transport in our devices is limited by $W$ rather than impurity scattering. For some realizations of BZ fermions, their mean free path also exceeds 10~{\textmu}m, marginally smaller than $l$ for Dirac fermions (cf. Figs 1b and 1d). Supplementary Note 4 provides similar analysis for $\mu$ and $l$ at other values of $q$. 

We corroborate the existence of ballistic BZ fermions using so-called bend resistance geometry~\cite{mayorov2011,beenakker1991} sketched in Fig. 2a. The geometry allows one to detect if charge carriers can move ballistically over the entire channel width $W$, along straight trajectories connecting current and voltage contacts (see caption of Fig. 2a). If this is the case, ballistic transport gives rise to the negative sign of the bend resistance $R_b$(ref~\cite{beenakker1991}) in contrast to its conventional, positive sign for diffusive (ohmic) transport. As expected from very long $l$ of Dirac fermions, $R_b$ was negative in zero $B$ everywhere away from NPs and vHS (Fig. 2b), confirming further the high quality of our superlattice devices. Finite $B$ bend Dirac fermion trajectories and, as expected, $R_b$ rapidly reversed its sign~\cite{beenakker1991} with increasing $B$ (inset of Fig. 2b). Remarkably, our devices exhibited negative $R_b$ also in high $B = B_{p/q}$, thus revealing straight trajectories over distances of several {\textmu}m (Figs. 2c,d and Supplementary Note 5). For comparison, the corresponding map of $\rho_{xx}$ is provided in Supplementary Note 6, which shows that the measured longitudinal resistance always remained positive. The profound negative pockets in $R_b(V_g,B)$ appeared only around $B$ corresponding to $\phi/\phi_0 = 1/2$, $1/3$, $1/4$ and $1/5$. This is the regime where the existence of BZ fermions, experiencing zero $B_{eff}$, was previously inferred~\cite{krishnakumar2017,krishnakumar2018} from maxima in $\sigma_{xx}$ and zeros in $\sigma_{xy}$. The possibility of ballistic transport of BZ fermions was also suggested using numerical simulations~\cite{diez2014}. The observed negative $R_b$ prove the previous conjectures unequivocally. The negative pockets in Fig. 2c are located between NPs and vHS for BZ fermions, similar to the case of Dirac fermions. Away from $B = B_{p/q}$, non-zero $B_{eff}$ bends the BZ fermion trajectories and the negative signal disappeared (inset of Fig. 2d), again like in the case of other Bloch quasiparticles (electrons and Dirac fermions). Our devices exhibited long-range ballistic transport of BZ fermions only for unit fractions, $\phi_0/q$. For example, Fig. 2c reveals a pronounced set of negative $R_b$ pockets at $\phi/\phi_0 = 1/5$ but no negative signal was observed for $2/5$ and $3/5$. This behavior probably stems from lower $\mu$ of BZ fermions with $p > 1$, which can be attributed to their larger effective masses~\cite{krishnakumar2018}. 

\begin{figure*}
    \centering
    \includegraphics[width=.95\textwidth]{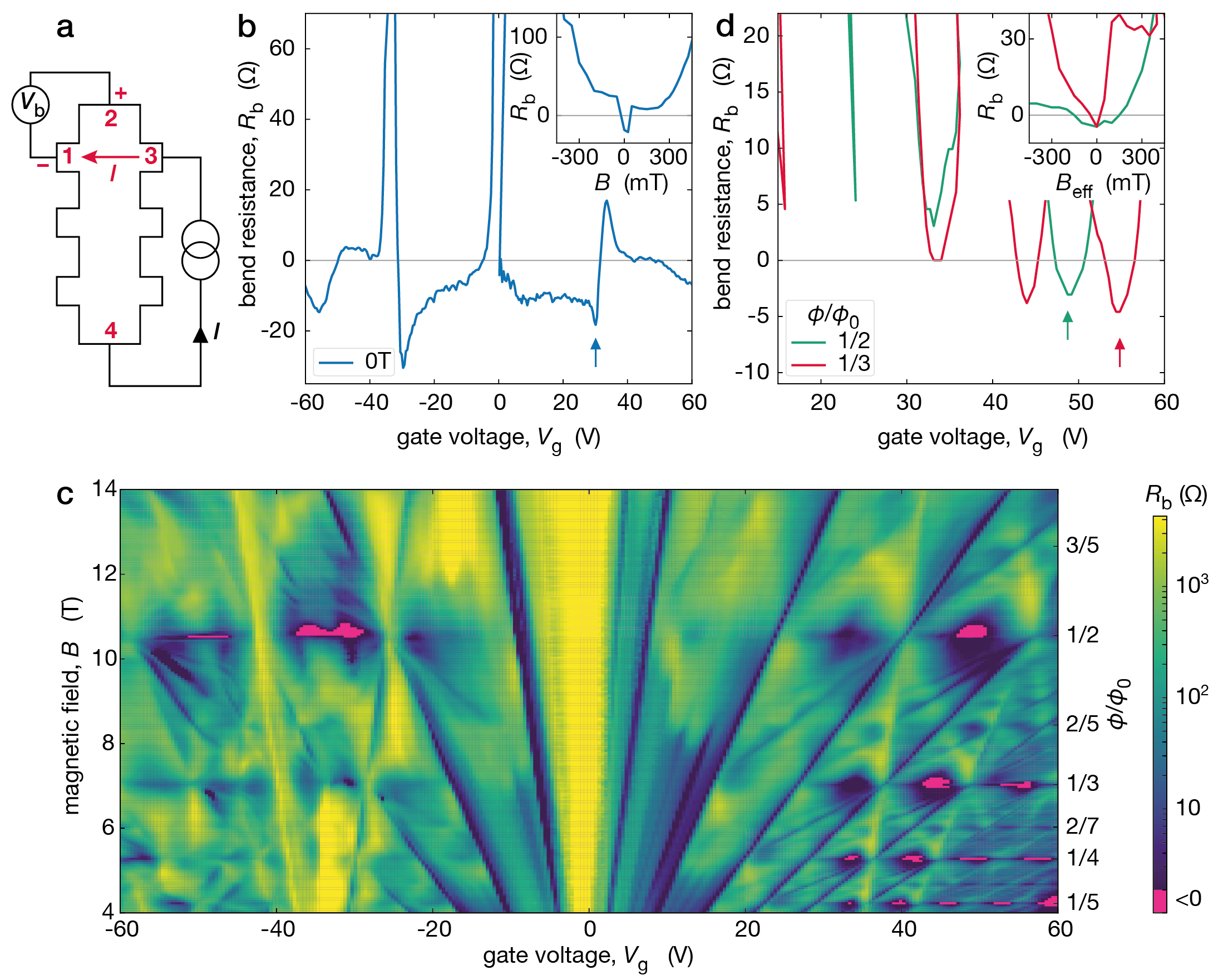}
    \caption{\textbf{Ballistic transport of BZ fermions over micrometer distances. a,} Schematic of bend resistance measurements. Current $I$ is applied between contacts 3 and 4, and voltage $V_b$ is measured between 2 and 1, yielding the bend resistance, $R_b$ = $V_b/I$. The voltage is positive for diffusive transport but becomes negative, if charge carriers move directly from current injecting contact 3 into voltage probe 1 (as shown by the red arrow). \textbf{b,} Bend resistance for Dirac fermions in zero $B$ (device D2 with $W$ = 4~{\textmu}m and $\theta \approx$ 0$^{\circ}$). Inset: $R_b(B)$ taken at the minimum indicated by the arrow in the main plot. \textbf{c,} Map $R_b(V_g,B)$ for the same device. $B$ was changed in steps of 50~mT. Pockets of negative $R_b$ appear along $\phi/\phi_0 = 1/q$ and are seen in magenta. \textbf{d,} Cross-sections from panel c for $q =$ 2 and 3. The inset shows sign reversals in $R_b$ plotted as a function of $B_{eff} = B – B_{p/q}$ for the minima marked by the color-coded arrows. $T$ = 2~K for all the plots.}
    \label{fig:fig2}
\end{figure*}

\subsection*{Lifting degeneracy in BZ minibands}
The high mobility of BZ fermions resulted in their Landau quantization in small $B_{eff} <$ 1~T, which allowed us to find experimentally the spectral degeneracy $g$ for different $p/q$. To this end, Fig. 3a shows a high-resolution map of $\sigma_{xx}$ between $\phi/\phi_0 = 1/2$ and $1/4$ (part of Fig. 1c). One can clearly see many Landau mini-fans originating from NPs for different realizations of BZ fermions. For example, there are three profound mini-fans spreading from $B \approx$ 9.7~T ($\phi/\phi_0 = 1/3$) for both negative and positive $B_{eff}$. For clarity, the conductivity map in Fig. 3a is replotted in Fig. 3b by tracing all well-defined minima in $\sigma_{xx}$. Each minimum can be described by its integer filling factor $\nu$, which we calculated from the minimum’s slope. This representation of the Hofstadter spectrum, where LLs are plotted as a function of $n$ or $V_g$ rather than energy, is usually referred to as the Wannier diagram~\cite{ponomarenko2013,dean2013,hunt2013,yu2014,diez2014,wang2015,spanton2018,wannier1978,hofstadter1976}. In the diagram of Fig. 3b, one can identify LLs for BZ-fermion realizations at $q$ = 2, 3, 4, 5, 7, 8, 9 and 11, and for $p$ = 1, 2, 3, 4 and 5. The difference in $\nu$ between the nearest LLs yields directly their $g$. For example, all LLs found for $\phi/\phi_0= 1/2$ were separated by $\Delta\nu$ = 2 whereas those at $\phi/\phi_0= 1/3$ and $1/4$ by $\Delta\nu$ = 3 and 4, respectively (see Fig. 3b). Therefore, the observed degeneracies were equal to $q$. Note that the measured Hall conductance also exhibited quantized values in steps of $qe^2/h$ (Supplementary Note 7). Examining the Wannier diagram further, we find that $\Delta\nu$ was equal to $q$, independently of numerator $p$ and for all LLs stemming from the same $B_{p/q}$. This observation agrees with the theoretical expectation that there should be $q$ equivalent mini-valleys for each realization of BZ fermions~\cite{brown1964,zak1964,azbel1964,wannier1978,hofstadter1976,streda1982,rhim2012,rammal1985,delplace2010,chen2014,tknn1982,landau1980}. The agreement takes into account that both spin and valley degeneracies for BZ fermions were lifted by exchange interactions~\cite{yu2014}, as in the case of the main Dirac spectrum with its clearly lifted degeneracies (see the LLs marked in black in Fig. 3b). Therefore, the total degeneracy for BZ minibands in Fig. 3 was $g = 4q$. This is further supported by our measurements at a relatively high $T$ of 2~K that suppressed LLs with lifted spin and valley degeneracies (Supplementary Note 6). The same behavior (Landau mini-fans exhibiting the sequence $\Delta\nu = q$) was also observed in other parts of the Wannier diagram for both electron and hole doping (see, e.g., Fig. 4). 

\begin{figure*}
    \centering
    \includegraphics[width=.95\textwidth]{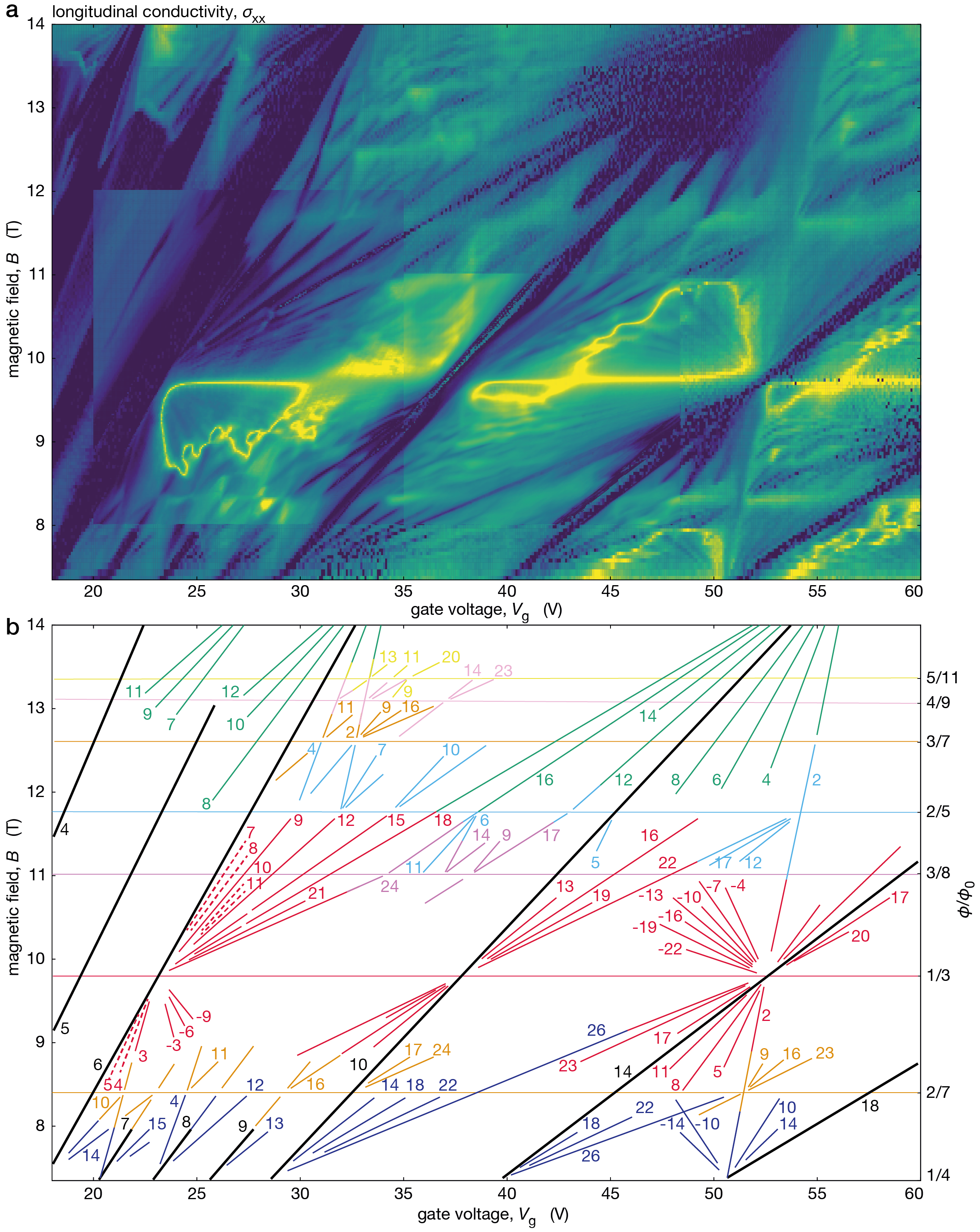}
    \caption{\textbf{Landau quantization in BZ minibands at 10 mK. a,} High-resolution map $\sigma_{xx}(V_g, B)$ for the electron-doped region indicated in Fig. 1c by the white rectangle (device D1). To better resolve LLs around $q = 3$, $B$ was changed in steps of 10 to 20~mT whereas data in other regions were acquired using 40 to 80~mT steps. This has resulted in the contrast discontinuities seen in the map. Log color scale: indigo (230~nS) to yellow (7.8~mS) for the entire map. \textbf{b,} Minima from panel (a) are shown schematically. The color-coded numbers are the filling factors for the corresponding LLs. Thick black lines correspond to the main sequence of LLs for graphene’s Dirac spectrum (spin and valley degeneracy lifted). The green, red, navy, blue, orange, magenta, pink and yellow lines correspond to $q =$ 2, 3, 4, 5, 7, 8, 9 and 11, respectively. Dashed red lines: minima due to lifted mini-valley degeneracy.}
    \label{fig:fig3}
\end{figure*}

There is however one notable exception, which was observed for $\phi/\phi_0 = 1/3$ (left part of the Wannier diagram in Fig. 3b; dashed lines). In this case, LLs of BZ fermions with $q = 3$ are separated only by $\Delta\nu = 1$ so that all consecutive LLs from 3 to 12 are present on the fan diagram. The indicated minima in $\sigma_{xx}$ were rather fragile, being rapidly smeared by $T$ or excitation current (Supplementary Note 7). Landau levels with $\Delta\nu$ = 1 at $\phi/\phi_0 = 1/q$ do not exist within the single-particle Hofstadter-Wannier model~\cite{wang2015,spanton2018,wannier1978,hofstadter1976}. Such \emph{extra} LLs have been observed previously and referred to as the fractional Bloch quantum Hall effect (FBQHE)~\cite{wang2015} or symmetry-broken Chern insulator (SBCI)~\cite{spanton2018}. Theoretically, the many-body states can arise from interaction effects that lift the mini-valley degeneracy of BZ fermions. This may happen via mini-valley mixing due to the formation of charge-density waves (commensurate with the magnetic superlattice composed of $q$ unit cells of the underlying moiré pattern) or a Wigner crystal~\cite{dasilva2016} (states localized in a part of the magnetic supercell). Alternatively, the fully lifted degeneracy may occur through spontaneous mini-valley polarization of BZ LLs, analogous to spin/valley ferromagnetism in graphene
~\cite{yu2014}. In the latter case, the BZ fermion states localize in the momentum space around one of the mini-valleys but remain delocalized across the magnetic supercell. 

\section*{Discussion}

\begin{figure*}
    \centering
    \includegraphics[width=.95\textwidth]{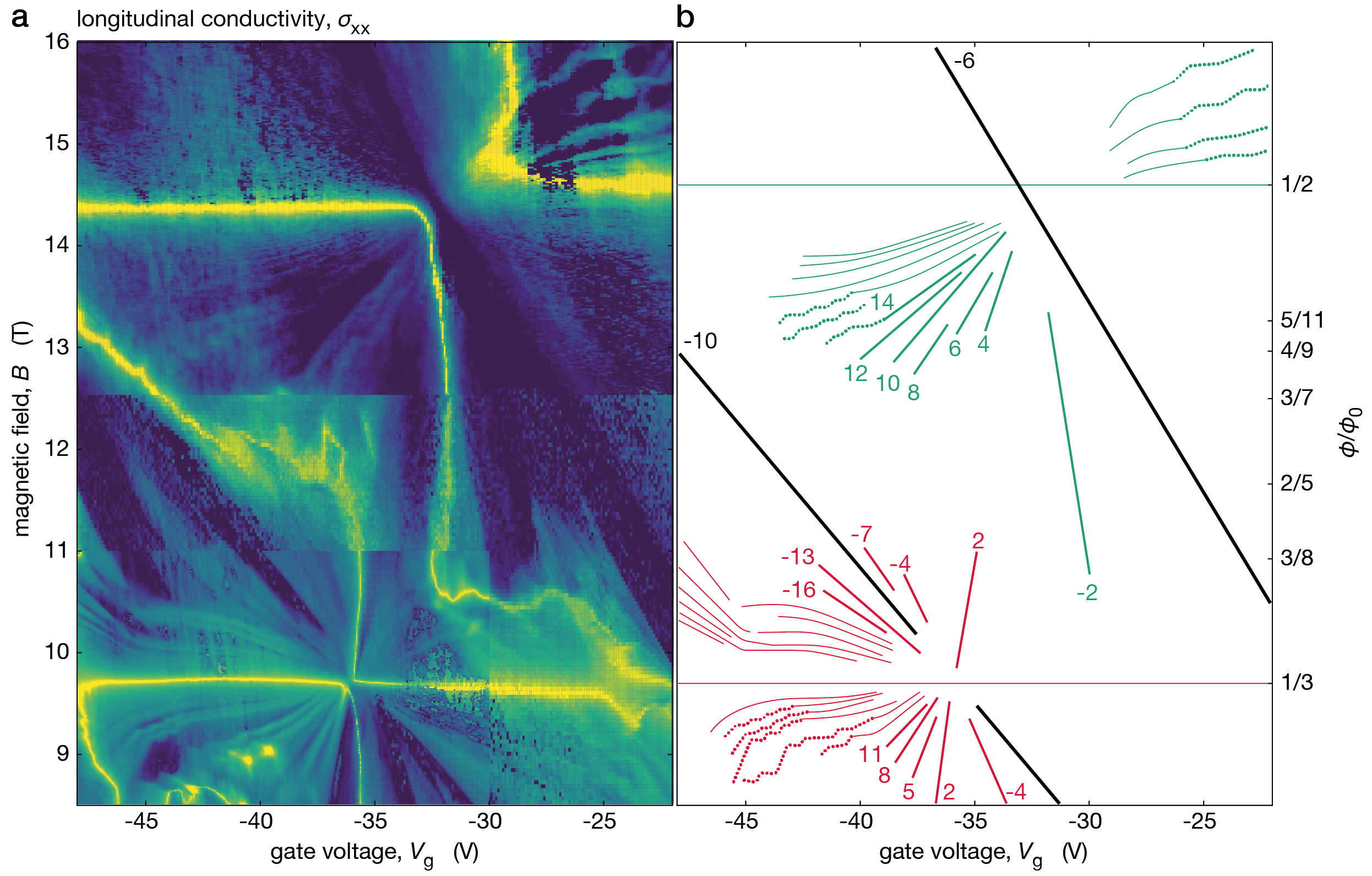}
    \caption{\textbf{Anomalous behavior of Landau levels in BZ minibands. a,} High-resolution $\sigma_{xx}(V_g, B)$ for the hole-doped region marked in Fig. 1c (device D1). The measurements below $B =$ 12.5~T were carried out at 10~mK, and at 250~mK in the fields above. $B$ was applied in steps of 16 and 40~mT above and below 12.5~T, respectively, except for the low left corner where the data were acquired at higher resolution of 10~mT (seen as discontinuities in the contrast). Log scale: indigo (310~nS) to yellow (0.3~mS). \textbf{b,} Schematics for the conductance minima found in (a). The same color coding as in Fig. 3b. The solid lines indicate LLs evolving as expected, linearly in $B$ and $V_g$. The thin curves indicate the anomalous bending whereas the dotted curves show the staircase-like evolution observed for some LLs. All the anomalous features were highly reproducible and, for example, did not depend on the step size in $B$. }
    \label{fig:fig4}
\end{figure*}

Some Landau mini-fans exhibited highly anomalous behavior at low $T$, which cannot be understood either within the Hofstadter-Wannier model or by considering LLs of non-interacting BZ fermions and, to the best of our knowledge, was never reported before. From either perspective, individual LLs must evolve linearly in $V_g$ and $B$, as indeed seen in most cases (e.g., Fig. 3). This is because the density of states on each BZ-fermion LL is proportional to $B_{eff} = B - B_{p/q}$. Also, the long ($\approx$300~nm) distance from graphene to the gate in our devices makes quantum capacitance corrections~\cite{chen2008,Yu2013} negligible, leading to the $n \propto V_g$ dependence. Unexpectedly, we found that some Landau fans of BZ fermions exhibited bending and staircase-like features. Figure 4 shows examples of such behavior for the case of hole doping. Both bending and staircases appear away from NPs, in the regions close to BZ fermions’ vHS. The observed bending towards the gate-voltage axis indicates that, in addition to the visible LLs, there can be some other electronic states that are populated in parallel. Because electron-electron interactions clearly play a significant role under the reported experimental conditions (e.g., they lift all the spectral degeneracies), it is reasonable to expect that interactions are also involved in the described anomalies. However, the usual suspects, such as negative compressibility cannot possibly explain our findings (in the latter case, LLs would bend toward the $B$ axis). The anomalous behavior is possibly due to an interplay of BZ-fermion LLs with quantized states originating from nearby vHS, which would lead to redistribution of charge carriers between states with the light and heavy effective masses. One possible scenario is localization of electrons within some parts of the magnetic supercell, because its size becomes notably larger than the magnetic length at $B >$ 10~T (Wigner crystallization mentioned above). Unfortunately, we could not find any additional features that would enable us to decipher origins of the described anomalies and, therefore, have to leave them for further investigation.

\paragraph{Methods}
The reported graphene-on-hBN superlattices were assembled using the standard dry transfer procedures and polydimethylsiloxane (PDMS) stamps coated with a polypropylcarbonate (PPC) layer7. The thicknesses of the used hBN crystals was between 20 and 70~nm. After the trilayer stack was assembled, we used the standard electron beam lithography and reactive ion etching to define trenches for electrical contacts. Cr/Au films were evaporated into the trenches. Finally, using ion etching again, the trilayer stack was shaped into a Hall bar mesa.
For electrical measurements, the standard low-frequency lock-in technique was employed. At temperatures below 100~mK, we used excitation currents of $\approx$10~nA, and between 0.1 to 1~{\textmu}A at higher $T$. The carrier concentration was controlled by applying a DC gate voltage between graphene and the silicon substrate. 

{\small{
\bibliographystyle{naturemag}
\bibliography{refs}
}}
	
\paragraph{Acknowledgements}
This work was supported by the Engineering and Physical Sciences Research Council (EPSRC), Lloyd’s Register Foundation, Graphene Flagship, the Royal Society and the European Research Council (grant Vander). A.I.B. acknowledges support from the Graphene NOWNANO Doctoral Training Centre.

\paragraph{Author contributions}
P.K., N.X., and M.H. fabricated devices. J.B., R.K.K., L.A.P., C.M., M.K., A.I.B. performed transport measurements with help from A.M, M.D.T and J.R.P. J.B., P.K., R.K.K., L.A.P., V.I.F, A.K.G. and A.I.B. performed data analysis. T.T. and K.W. provided the hBN crystals. J.B., P.K., V.I.F., A.I.B. and A.K.G. wrote the manuscript. All authors contributed to the discussions.

\paragraph{Competing interests}
The authors declare no competing interests

\end{multicols}
\end{document}


\setcounter{page}{10}
\beginsupplement
\section*{Supplementary information}
\paragraph{Supplementary note 1: Studied devices.}
We studied six different superlattice devices as summarized in Supplementary Table 1. Here $W$ refers to the device width, $C_g$ is its gate capacitance and $\theta$ is the angle between the crystallographic axes of hBN and graphene. The latter was calculated from the measured period of BZ oscillations~\cite{krishnakumar2017si}.

\begin{table}[h]
 \centering
 \caption{
 }
 \begin{tabular}{lrrr}
  \textbf{Device} & $\mathbf{W}$ \textbf{({\textmu}m)} & $\boldsymbol{\theta}$ \textbf{($^{\mathbf\circ}$)} & $\mathbf{C_g}$ \textbf{({\textmu}F/m$^2$)} \\
  \hline\\
  D1 & 17 & 0.4 & 101 \\
  D2 & 4 & 0 & 98 \\
  D3 & 3 & 0.2 & 104 \\
  D4 & 3.2 & 0.5 & 96 \\
  D5 & 2 & 0.2 & 106 \\
  D6 & 2 & 0.3 & 98 \\
 \end{tabular}
 \label{tab:si1}
\end{table}
	
\paragraph{Supplementary note 2: Magnetic focusing in graphene superlattices.}
As a further confirmation of the devices’ high quality, we report transverse magnetic focusing (TMF) experiments at low $B$ (Supplementary Fig. 1). The observation of resistance oscillations due to TMF confirms that Dirac fermions travel ballistically across the device~\cite{lee2016si,berdyugin2020si}, forming skipping orbits extending over hundreds of superlattice unit cells. TMF measurements are also known to provide information about the Fermi surface topography in clean metals, including graphene superlattices~\cite{lee2016si,berdyugin2020si}. Our TMF results are in agreement with those reported previously~\cite{lee2016si,kraft2020si}.
 
\begin{figure}[h]
 \centering
 \includegraphics[width=.55\textwidth]{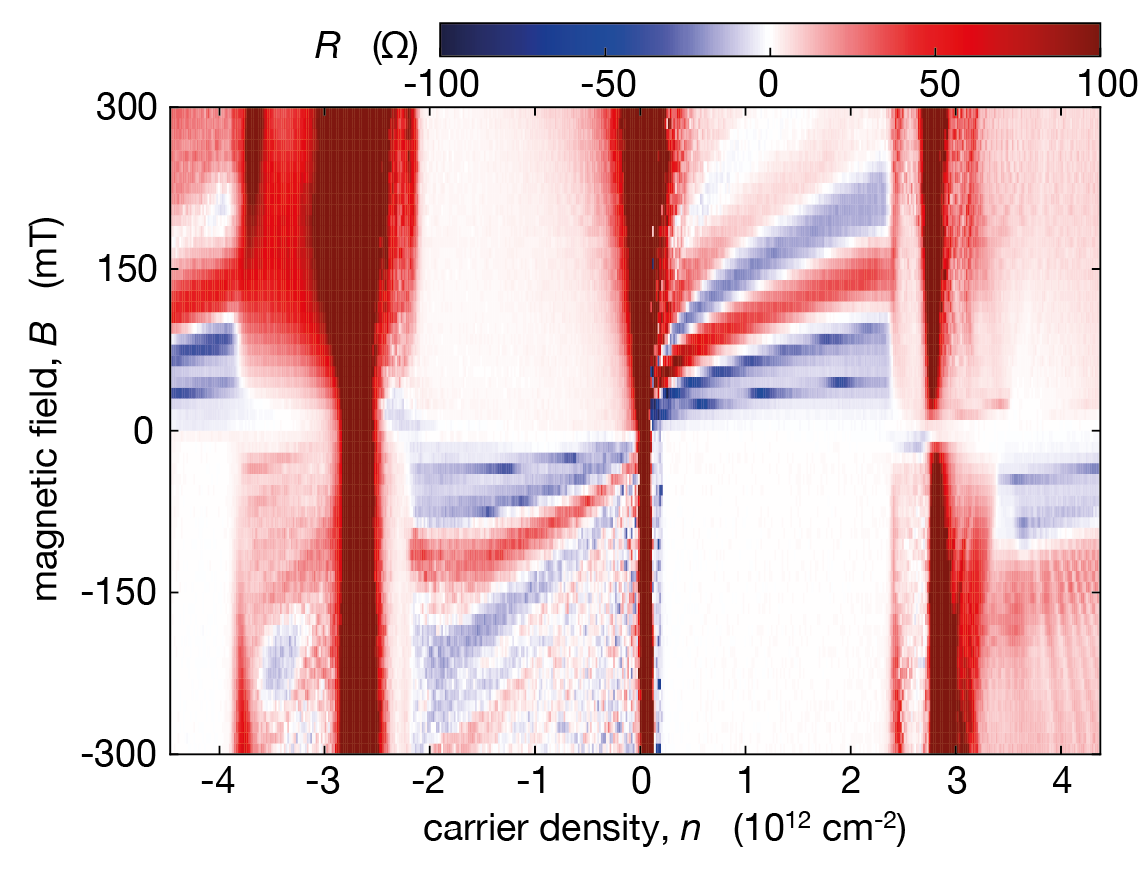}
 \caption{\textbf{Transverse magnetic focusing.} Measurements using contacts separated by 1.5~{\textmu}m (closest contacts for device D1 pictured in Fig. 1a of the main text); $T$ = 10~mK.}
 \label{fig:si1}
\end{figure}
\newpage
\paragraph{Supplementary note 3: Determining mobility and mean free path of BZ fermions.}
To evaluate the mobility of Brown-Zak fermions, we use the standard formula $\mu=\sigma_{xx}/(n_{BZF} e)$ where $n_{BZF}$ is the carrier density of BZ fermions and $\sigma_{xx}=1/\rho_{xx}$ . Note that the latter expression is exact at $\phi=\phi_0 p/q$ (that is, it does not contain $\rho_{xy}$ because the effective magnetic field $B_{eff}$ acting on BZ fermions is zero). To determine $n_{BZF}$ for a given $V_g$, we first used Hall measurements at small fields $B \leq$ 0.1~T to determine the geometrical capacitance. Then, using longitudinal conductivity maps around the $p/q$ fractions, we identified positions of the neutrality points (NPs) as $V_g$ into which Landau mini-fans converged (see Fig. 1c and Fig. 3a of the main text). Finally, vHS were identified from Hall effect measurements as $V_g$ where $\rho_{xy}$ changed its sign without exhibiting mini-fans (Supplementary Fig. 2b). As $n_{BZF}$ varies linearly across NPs and exhibits jumps at vHS, the known geometrical capacitance allowed us to reconstruct $n_{BZF}(V_g)$ as shown in Supplementary Fig. 2a.
The mean free path $l$ was calculated using the standard formula $\sigma_{xx}=g e^2/h((k_F l)/2)$ where the Fermi wave vector $k_F=(4\pi n_{BZF}/g^(1/2)$ also depends on the BZ fermion degeneracy $g$. The final expression reads 

\begin{equation}
l= \frac{2}{\rho_{xx}} \frac{\hbar}{e^2} \sqrt{\frac{\pi}{g~n_{BZF}}}
\end{equation}

\begin{figure}[h]
 \centering
 \includegraphics[width=.8\textwidth]{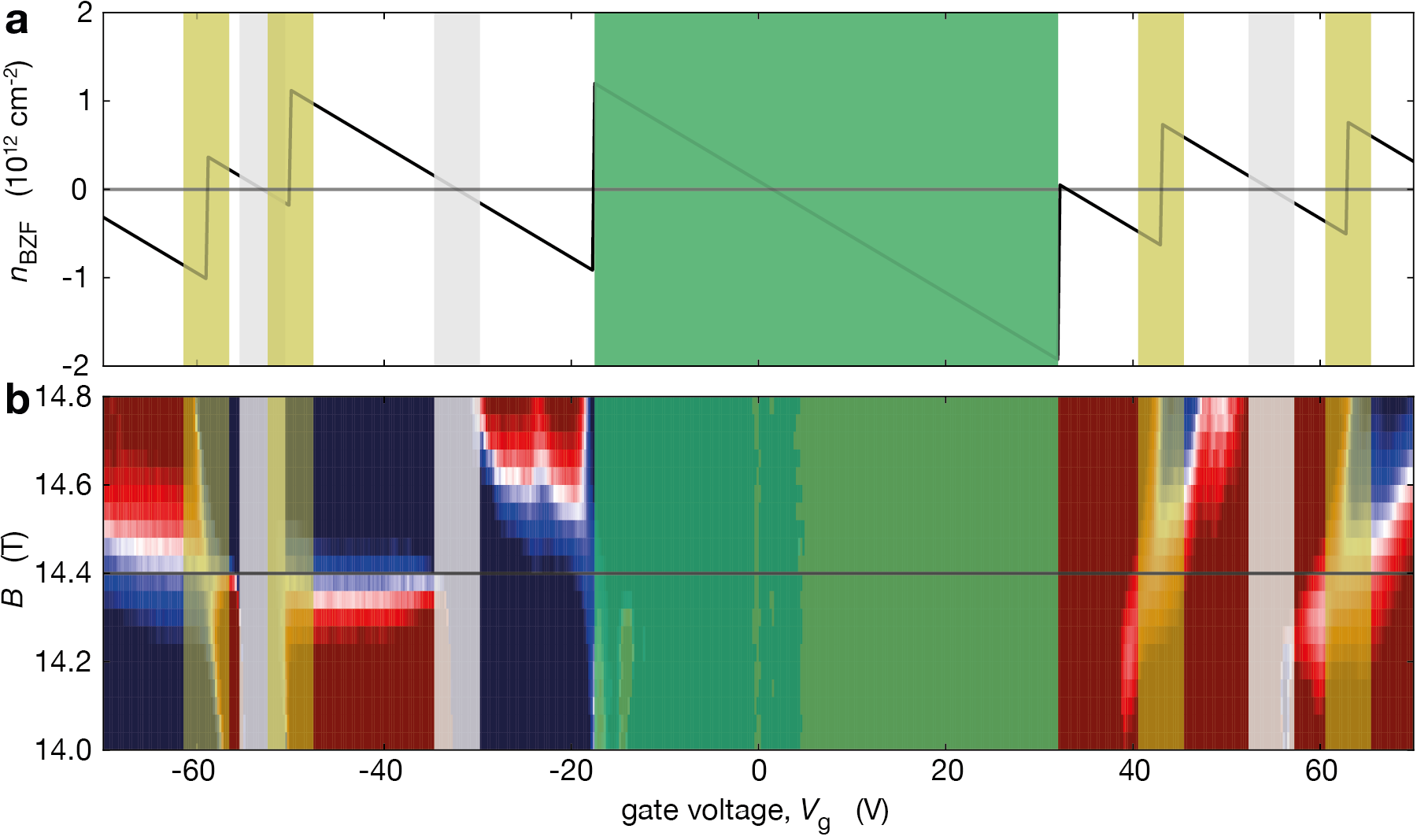}
 \caption{\textbf{Evaluating density of BZ fermions.} Dependence of $n_{BZF}$ on gate voltage at $\phi/(\phi_0=1/2)$ for device D1. b, Measured maps for the Hall resistivity around $\phi/(\phi_0=1/2)$. Colour scheme: blue and red represent negative and positive $\rho_{xy}$, respectively. Regions around NPs are indicated by the grey semi-transparent strips. The yellow strips mark vHS. The central green area covers the region dominated by the quantum Hall effect of Dirac fermions from the main graphene spectrum (see Fig. 1c of the main text).}
 \label{fig:si2}
\end{figure}

\begin{figure}
 \centering
 \includegraphics[width=.9\textwidth]{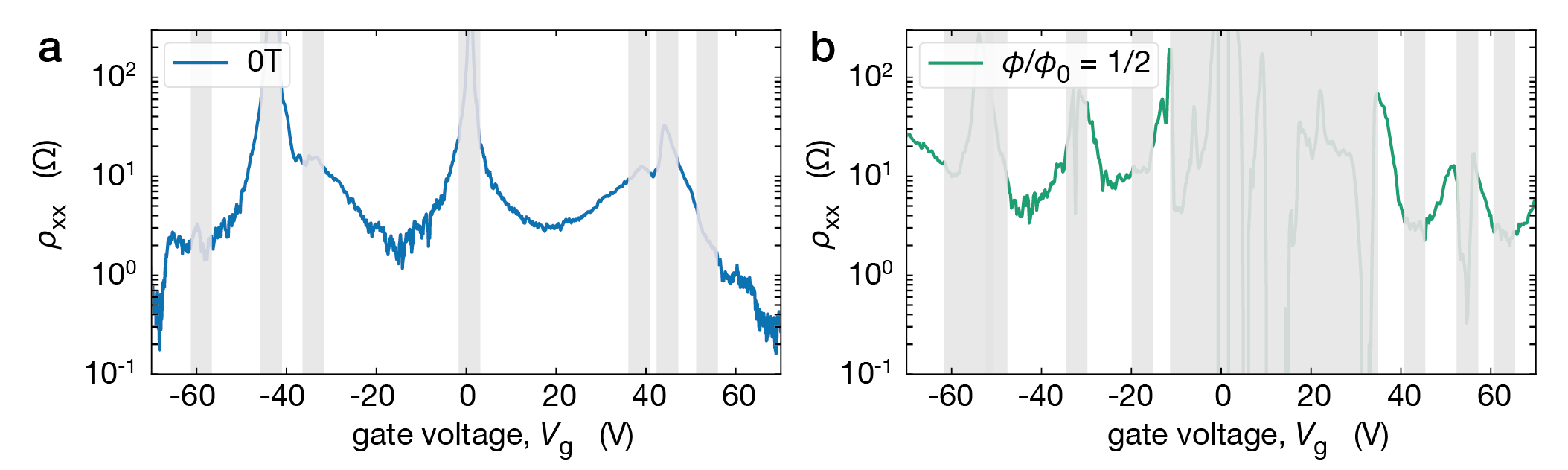}
 \caption{\textbf{Longitudinal resistivity $\boldsymbol{\rho}_{\mathbf{xx}}$ for device D1.} a, Measurements in zero field and b, for $\phi/\phi_0 =1/2$. These data were used to calculate the mobilities and mean free paths in Fig. 1 of the main text.\\[3ex]}
 \label{fig:si3}
\end{figure}

\newpage

\paragraph{Supplementary Note 4: BZ fermions at higher order fractions.} In Fig. 1 of the main text, we presented $\mu$ and $l$ for Dirac fermions and for BZ fermions at $\phi/\phi_0 =1/2$. For completeness, Supplementary Fig. 4 shows the same analysis for the case of $\phi/\phi_0 = 1/3$ and $1/4$. One can see that mobilities of BZ fermions with the larger $q$ still remain of the order of 10$^6$~cm$^2$/Vs and their mean free path approaches values comparable to the device width $W$, which suggests a notable contribution from edge scattering. 
	
\begin{figure}[h!]
 \centering
 \includegraphics[width=.9\textwidth]{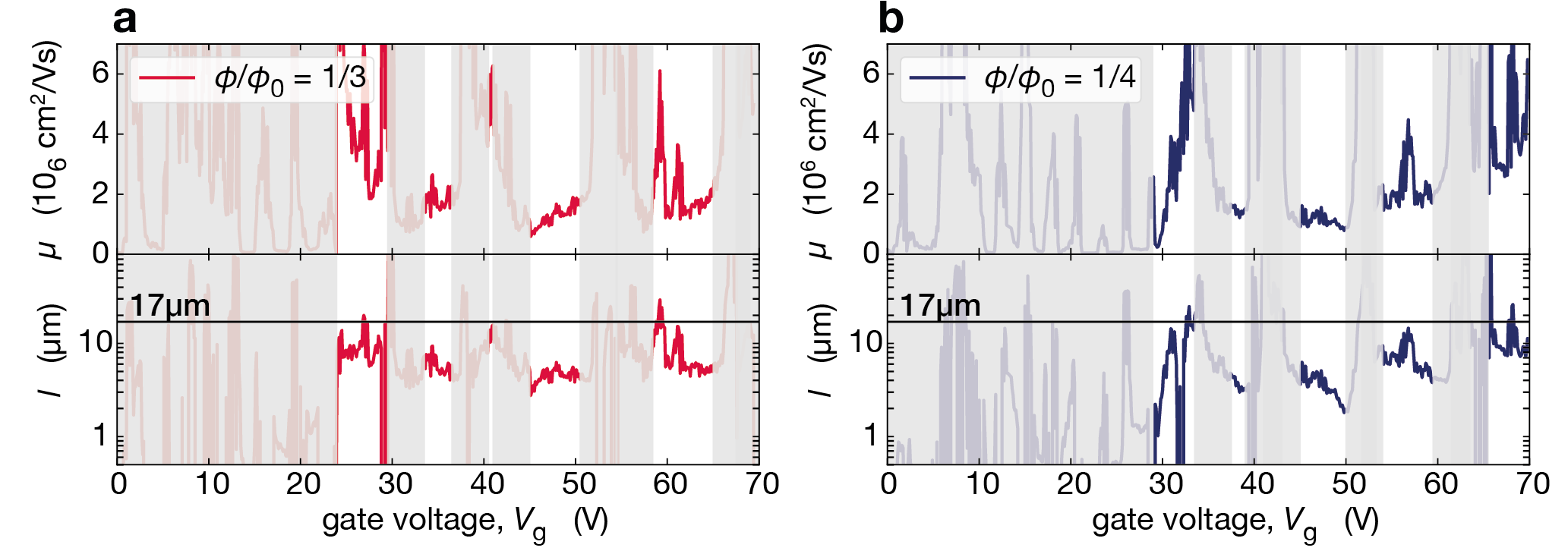}
 \caption{\textbf{Ballistic transport of BZ fermions at unit fractions of the flux quantum.} The data are for device D1 at 10~mK for $\phi/\phi_0 =1/3$ (a) and $1/4$ (b). The same presentation as in Figs. 1b,d of the main text. We show the data for positive voltages because for mini-fans and vHS could accurately be identified only for electron doping.}
 \label{fig:si4}
\end{figure}

\newpage

\paragraph{Supplementary Note 5: Additional examples of ballistic transfer of BZ fermions.} In the main text, we have emphasized that, at fields $B = B_{p/q}$, BZ fermions move through the superlattice as if the applied field were zero. The effective mass of BZ fermions depends on $p/q$ because electronic spectra differ in different magnetic minibands. Away from the exact $B_{p/q}$ values, BZ fermions are expected~\cite{krishnakumar2017si} to experience an effective magnetic field $B_{eff} = B – B_{p/q}$ and, therefore, replicate magneto-transport effects known for charge carriers in conventional 2D electronic systems. This includes the negative bend resistance that is one of the most distinct, qualitative signatures of ballistic transport of charge carriers~\cite{beenakker1989si,beenakker1991si,gilbertson2011si}. The effect can be understood as follows. With reference to Fig. 2a of the main text, let us for simplicity consider positive charge carriers (hole-doping regime). If holes injected from contact 3 can travel ballistically over a distance exceeding $W$ (that is, can reach contact 1 without scattering), an extra positive charge would be accumulated near contact 1. As a result, the voltage difference $V_{21}=V_2-V_1$ should be negative (see Fig. 2a of the main text). In contrast, if the transport is conventional (diffusive), holes from contact 3 travel along lines of the electric field and accumulate at contact 4. Accordingly, the sign of $V_{21}$ should be conventional (that is, positive). The same consideration for $V_{21}$ is valid for electrons. Therefore, the negative sign of $R_b$ signifies ballistic transport over distances larger than $W$. 
Negative $R_b$ was reported in Fig. 2 of the main text for one of our devices (D2). Supplementary Fig. 5 provides further examples of ballistic transport of BZ fermions using two other superlattices (devices D3 and D4). Pronounced pockets of negative $R_b$ are seen in Supplementary Fig. 5 at unit fractions of $\phi_0$ with $q$ from 2 to 5. Despite relatively small $W \approx$ 3~{\textmu}m, no evidence for ballistic transfer was observed for high-order BZ states ($p >$ 1), in agreement with the results reported in the main text.
Note that occasionally we observed negative bend resistance away from $\phi/\phi_0 = 1/q$ (see, e.g., the vertical magenta stripe close to zero $V_g$ in Supplementary Fig. 5a). Unlike the ballistic transfer resistance at unit flux fractions, negative signals away from the unit fractions were not reproducible in different contact configurations. Such \emph{extra} negative signals are not surprising in our experimental geometry and well known to appear in the quantum Hall effect regime using narrow (mesoscopic) devices~\cite{buttiker1988si}.

\begin{figure}[h]
 \centering
 \includegraphics[width=.7\textwidth]{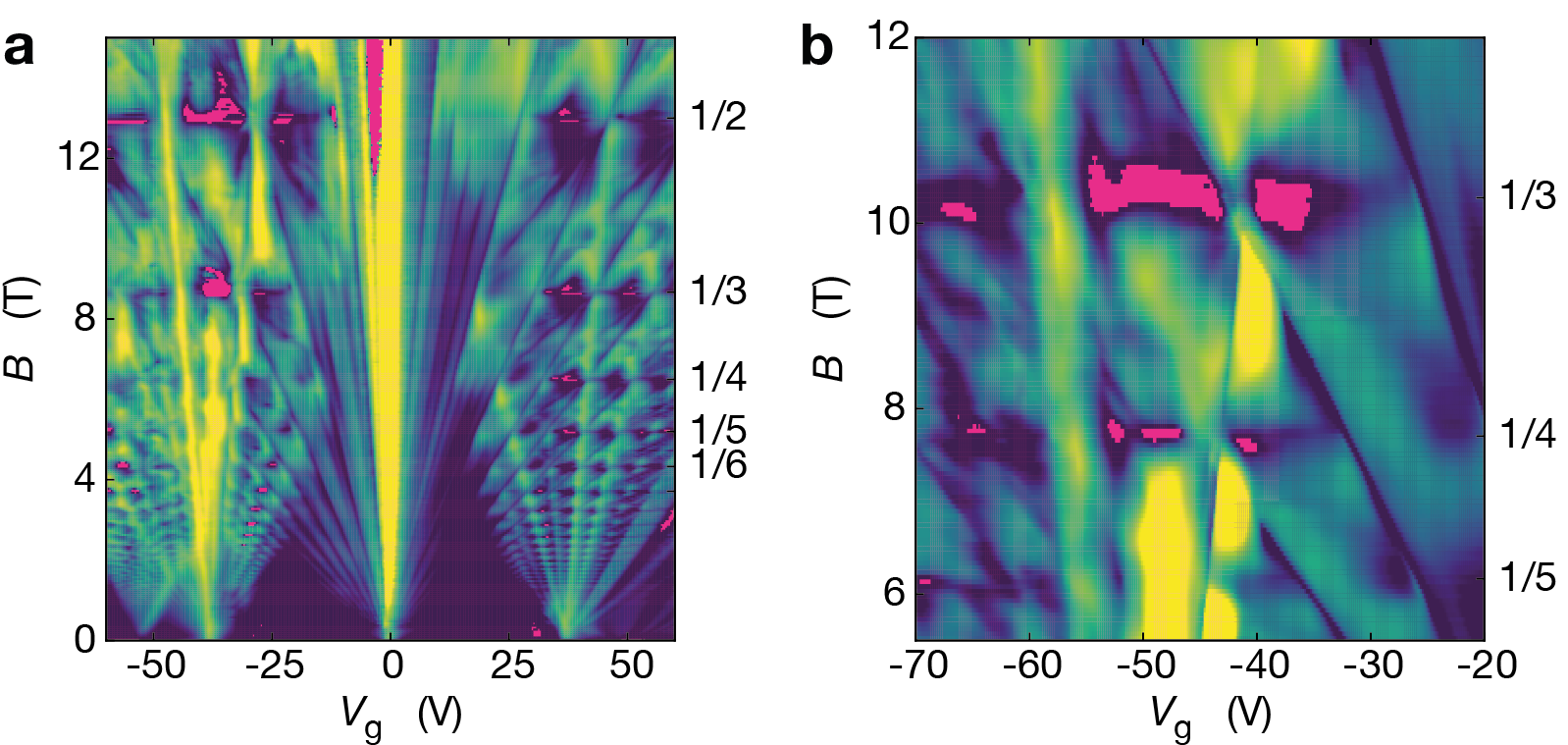}
 \caption{\textbf{Ballistic transport of BZ fermions over micrometer distances.} Fan diagrams obtained in the bend resistance geometry for devices D3 (a) and D4 (b) with $W$ = 3 and 3.2~{\textmu}m, respectively. $T$ = 2~K. Pockets of negative $R_b$ are highlighted in magenta. Indigo-to-yellow: Log scale truncated between 10 and 2,000~{\textOmega} to optimize the contrast.}
 \label{fig:si5}
\end{figure}

Ballistic transport of BZ fermions was found to be rather sensitive to $T$, and the pockets of negative $R_b$ universally disappeared above 30-50~K as shown in Supplementary Fig. 6. This is generally expected because the mean free path of BZ fermions should become shorter at higher $T$. However, the exact scattering mechanism could be nontrivial (see, e.g. Umklapp electron-electron scattering~\cite{wallbank2019si}) and requires further investigation. 
	
\begin{figure}[h]
 \centering
 \includegraphics[width=.5\textwidth]{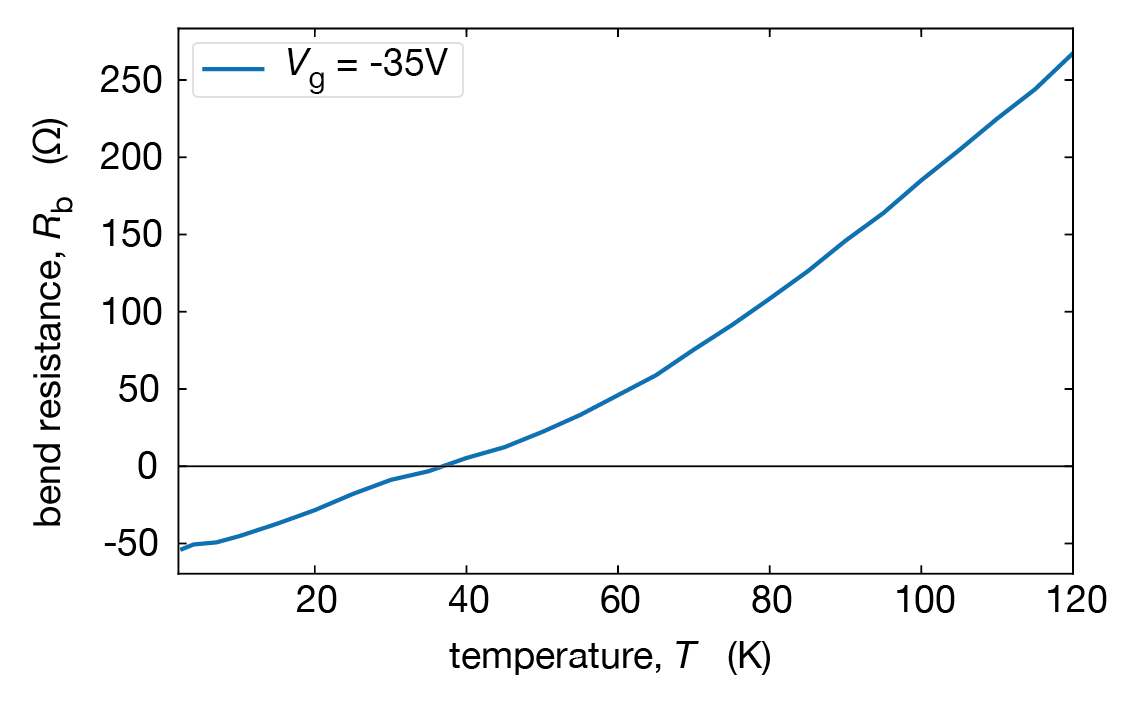}
 \caption{\textbf{Temperature dependence of BZ fermions’ ballistic transport.} An example of the bend resistance measured at $\phi/\phi_0=1/2$ using device D5 with $W$ = 2~{\textmu}m.\\[2ex]}
 \label{fig:si6}
\end{figure} 

\newpage
\paragraph{Supplementary Note 6: Supporting measurements in the longitudinal geometry.} To crosscheck our conclusions about ballistic transport of BZ fermions, we compare the negative bend resistance measurements shown in Figs. 2c,d of the main text with those made in the conventional longitudinal geometry for the same device D2 (Supplementary Fig. 7a). The longitudinal resistance $R_{xx}$ for BZ fermions was found positive in all the regions of the map where the negative bend resistance was reported, which corroborates the conclusion in the main text about ballistic transfer of BZ fermions across the device. 

\begin{figure}[h]
 \centering
 \includegraphics[width=.9\textwidth]{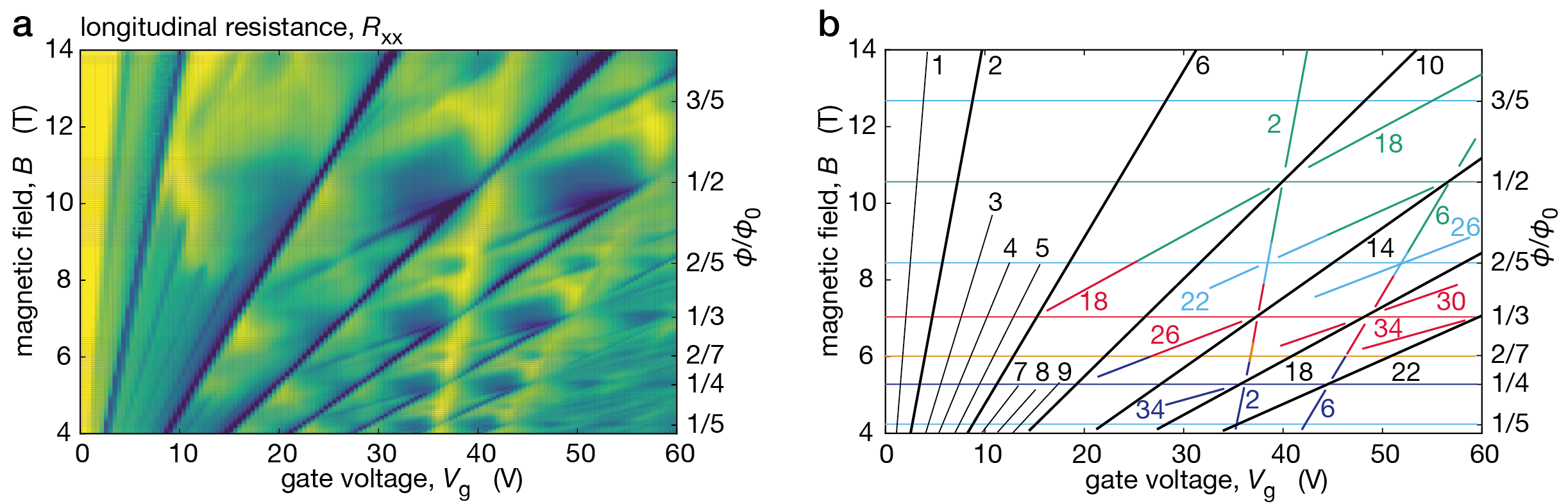}
 \caption{\textbf{Longitudinal resistance for ballistic BZ fermions.} a, $R_{xx}$ as a function of gate voltage and magnetic field measured at 2~K for device D2. Color scale is the same as in Fig. 2c of the main text. b, Minima found in the longitudinal conductivity are shown schematically. The color-coding is the same as for device D1 in Fig. 3b of the main text. The thin black lines mark LLs with the lifted spin and valley degeneracy for Dirac fermions of the main spectrum.}
 \label{fig:si7}
\end{figure} 

Let us note here that, according to the group theory of irreducible representations for the group of translations in a magnetic field, an electronic spectrum for each realization of BZ fermions should have an additional $q$-fold degeneracy. This is prescribed by the fact that a group corresponding to any $p/q$ fraction is non-Abelian (due to Aharonov-Bohm phases acquired upon translations in non-colinear directions) but contains an Abelian subgroup of translations corresponding to a magnetic superlattice with a $q$ times larger supercell. The additional $q$-fold degeneracy takes the form of $q$ mini-valleys in the magnetic mini Brillouin zone with an area $q$ times smaller than the moiré superlattice Brillouin zone at $B = 0$. This degeneracy is additional to the 4-fold spin and valley degeneracy of graphene’s original spectrum.

With this consideration in mind, the measurements in Supplementary Fig. 7 also support our other conclusion that the full degeneracy of BZ fermions is $4q$. Indeed, the $q$-fold degeneracy reported in Fig. 3 of the main text corresponds to the case where both spin and valley degeneracies of both Dirac and BZ fermions were lifted. Supplementary Fig. 7 shows LL fans at 2~K, the temperature much higher than 10~mK for the measurements in Fig. 3. Dirac fermions of the main spectrum exhibit the lifted spin and valley degeneracies by the relatively strong $B$ (thin black lines in Supplementary Fig. 7b). At lower fields $B <$ 3~T, these interaction-induced gaps become progressively smeared. As for BZ fermions, their mini-fans visible in Supplementary Fig. 7b reach only the effective field $|B_{eff}| <$ 2~T, which does not allow the lifting of spin and valley degeneracies at this temperature. Accordingly, only the main sequence of LLs for BZ fermions could be observed at 2~K, and it corresponds to the $4q$-fold degeneracy, as expected and explained in the previous paragraph.\\
	
\paragraph{Supplementary Note 7: Lifting mini-valley degeneracy.}
In the main text we reported additional quantum Hall effect minima that cannot be explained within the single-particle Hofstadter-Wannier (dashed lines in Fig 3b of the main text). Those minima in $\sigma_{xx}$ were attributed to BZ states with lifted mini-valley degeneracy. As an additional proof for the observed degeneracy lifting, Supplementary Fig. 8 shows measurements of Hall conductivity $\sigma_{xy}$ for the relevant range of $B$ and $V_g$ where the dashed lines occur in Fig. 3b. One can see well developed plateaus with the quantized values that are fully consistent with the filling factors reported in the main text and marked in Fig. 3b. This observation strongly supports our conclusions about lifting of all the degeneracies of BZ fermions at low $T$.
	
\begin{figure}
 \centering
 \includegraphics[width=.45\textwidth]{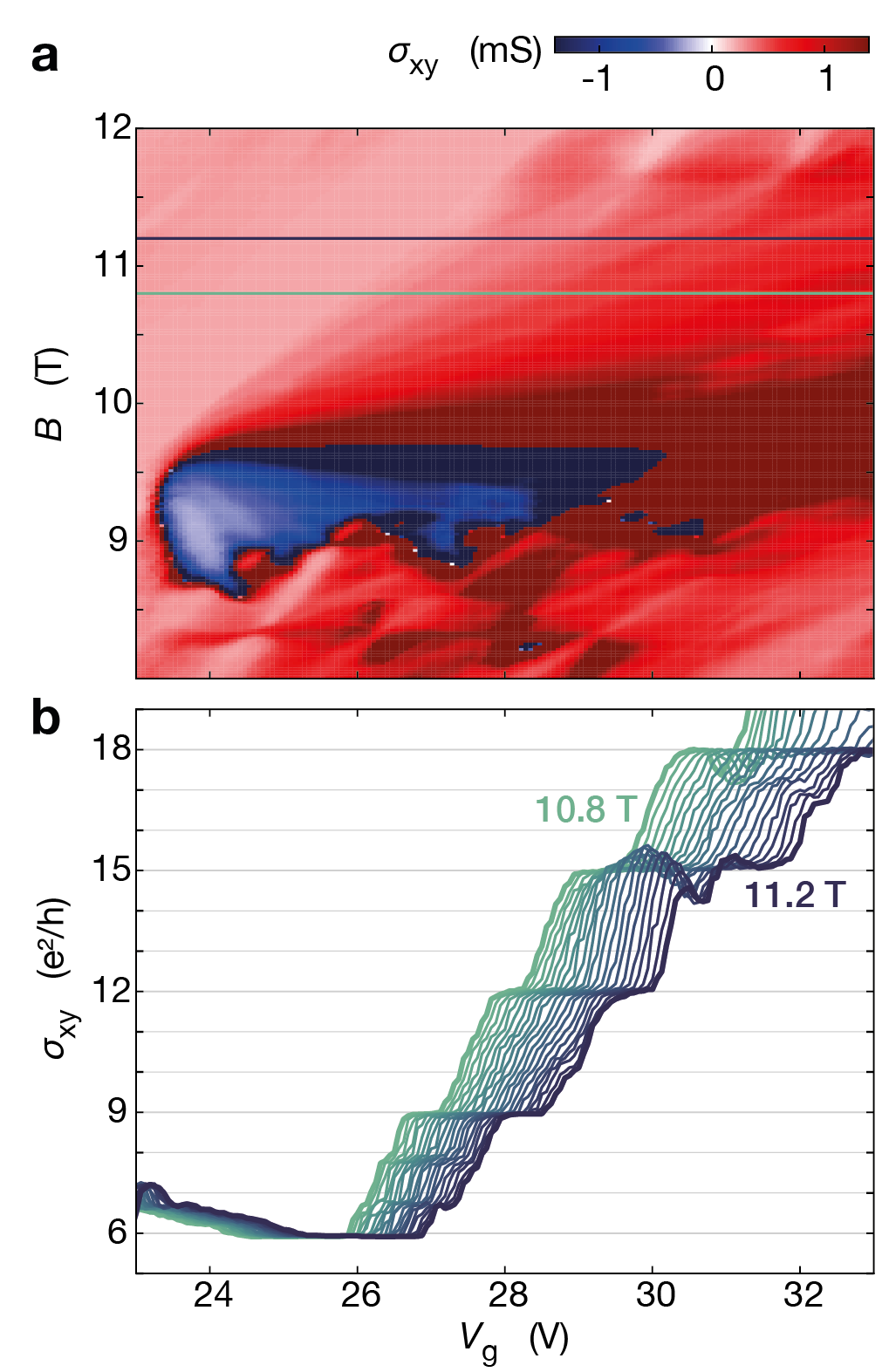}
 \caption{\textbf{Quantized Hall conductance for BZ fermions.} a, $\sigma_{xy}$ around $\phi/\phi_0 = 1/3$. b, Hall conductivity as a function of gate voltage at a number of constant $B$ within the field interval around 11~T (color-coded). The interval is marked by the horizontal lines in (a).}
 \label{fig:si8}
\end{figure} 
	
The described lifting of mini-valley degeneracy involves very small energy gaps as witnessed by rapid disappearance of the corresponding features with increasing $T$. Indeed, the quantized Hall plateaus seen in the above figure and the conductance minima marked by the dashed lines in Fig. 3b of the main text could not be resolved at 2~K. The features also disappeared rapidly with increasing the excitation current. For example, Supplementary Fig. 9 shows a Landau mini-fan around $\phi/\phi_0 = 1/3$ for currents of 10 and 100~nA. In the former case (Supplementary Fig. 9a), there are clear minima associated with to the lifted mini-valley degeneracy. The higher current (100~nA) resulted in complete smearing of these mini-gaps (Supplementary Fig. 9b), presumably because of an increase in the electronic temperature. 
	
\begin{figure}
 \centering
 \includegraphics[width=.8\textwidth]{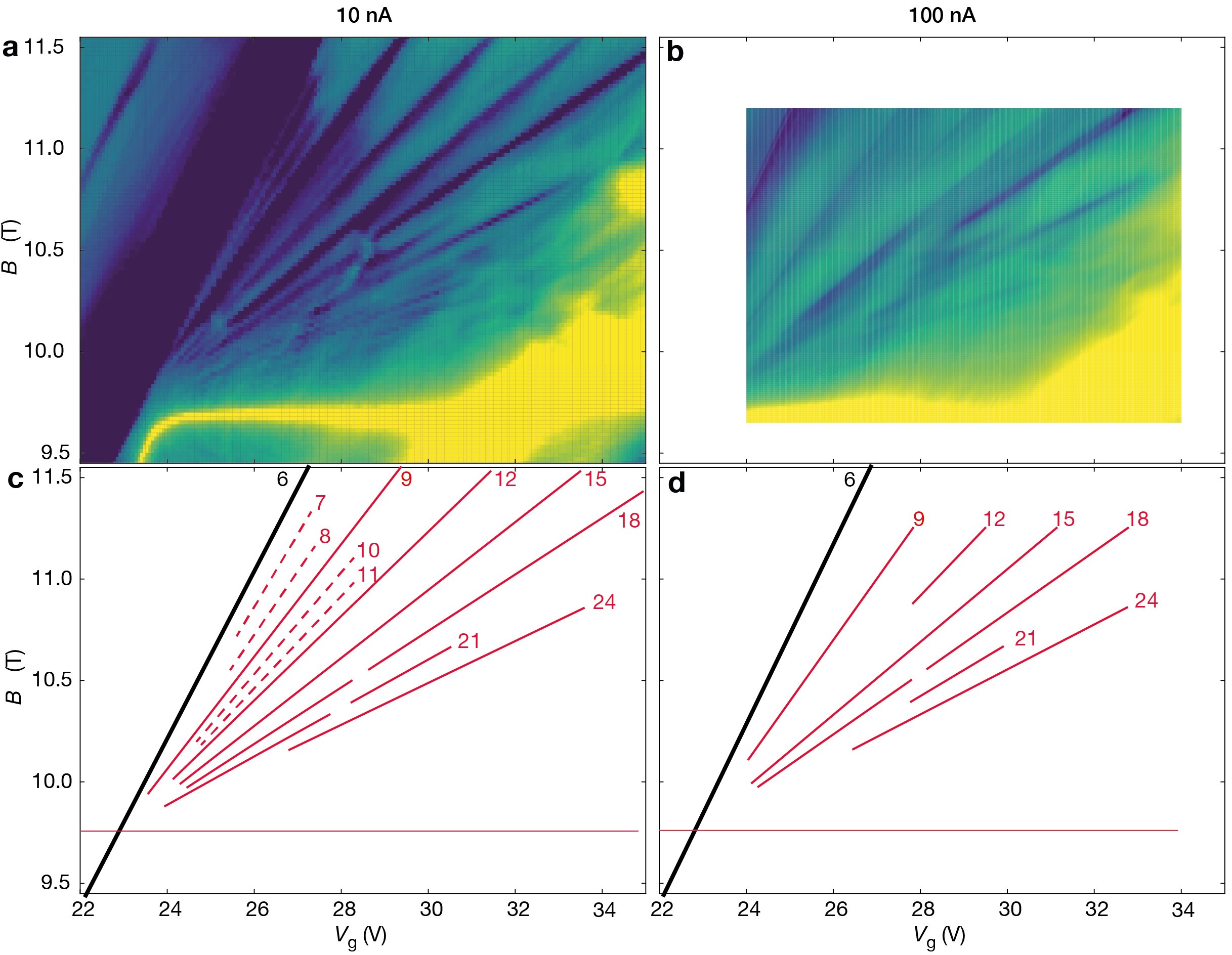}
 \caption{\textbf{Landau mini-fans for different excitation currents.} a and b, $\sigma_{xx}(B,V_g)$ at 10~mK for 10 and 100~nA, respectively. Indigo-to-yellow log scale: 310~nS to 780~{\textmu}S. c and d, Minima found in (a) and (b) are shown schematically. The color-coded numbers are the filling factors for the nearby LLs. Thick black lines: Main sequence of LLs for graphene’s Dirac spectrum.}
 \label{fig:si9}
\end{figure}

Finally, let us draw attention to the rather unusual re-entrant behavior seen for the mini-fan around $V_g$ = 28~V in Fig. 3 and Supplementary Fig. 9. The BZ-fermion gaps for $\nu$ = 18 and 21 seem to close within a certain interval $B$ and $V_g$. We attribute this closure to competition between these BZ states and the $\nu$ = 7 state from the main Dirac sequence. The likely mechanism of the suppression of exchange gaps is discussed in ref. \cite{yu2014si}.

\begin{multicols}{2}
\small{
\bibliographystyle{naturemag}
\bibliography{supplementary}
}
\end{multicols}